\documentclass[prd,showpacs,amsmath,showkeys,twocolumn,floatfix]{revtex4}
\usepackage[]{epsfig,dcolumn}
\usepackage{graphicx}
\usepackage{bm}

\def\beq{\begin{equation}}
\def\eeq{\end{equation}}
\def\beqa{\begin{eqnarray}}
\def\eeqa{\end{eqnarray}}
\def\psid{\psi^{\dag}}
\def\hn{{\hat n}}

\def\x{{\bf x}}
\def\y{{\bf y}}
\def\n{{\bf n}}

\def\k{{\bf k}}

\def\lsim{\mathrel{\rlap{\lower4pt\hbox{\hskip1pt$\sim$}}
    \raise1pt\hbox{$<$}}}
\def\gsim{\mathrel{\rlap{\lower4pt\hbox{\hskip1pt$\sim$}}
    \raise1pt\hbox{$>$}}}

\begin{document}

\def\ii{\'\i}

\title{
 An exactly solvable limit of low energy QCD }

\author{Peter O. Hess}
\affiliation{ Instituto de Ciencias Nucleares,  \\
Univiversidad Nacional Aut\'onoma de M\'exico,
Apdo. Postal 70-543, M\'exico 04510 D.F.}

\author{Adam P. Szczepaniak}
\affiliation{
  Department of Physics and Nuclear Theory Center \\
  Indiana University, 
  Bloomington, Indiana   47405-4202}

%\author{Peter O. Hess$^1$, Adam P. Szczepaniak$^2$ }
%\affiliation{1) Instituto de Ciencias Nucleares,  \\
%Univiversidad Nacional Aut\'onoma de M\'exico,
%Apdo. Postal 70-543, M\'exico 04510 D.F. \\
%2) Physics Depratment and Nuclear Theory Center, \\
%Indiana University,  Bloomington, Indiana 47405-4202
%}

\begin{abstract}
{Starting from the QCD Hamiltonian,
we derive a schematic Hamiltonian for low energy quark dynamics
with quarks  restricted to the lowest s-level. The resulting eigenvalue
problem can be solved analytically.
Even though the Hamiltonian
exhibits explicit chiral symmetry the severe restriction of the number of
degrees of freedom breaks the
pattern
of chiral symmetry breaking for finite quark masses.}

\end{abstract}
\maketitle

\section{Introduction}
As significant advances have been made over the past few years in
lattice gauge QCD \cite{lattice},  progress in developing other, approximate solutions to low energy QCD has not been as impressive.
Many models, such as the quark model, the bag model,
the flux tube model, and many others have been utilized to capture
selective feature of the theory. Traditional approaches to develop
models with a
more rigorous
relation to QCD are based on the covariant representation of the theory. Recently, however, fixed gauge approaches have also been intensively pursued  as they offer a bridge between QCD  and the more traditional (non-relativistic) many-body problems in nuclear  and condensed-matter physics. The disadvantage of a fixed gauge approach is, however, that it is not manifestly Lorentz invariant and thus there are fewer  restrictions on the dynamical operators resulting in general in complicated Hamiltonians.
Fortunately experience gained form studies of other
many-body systems can be helpful in identifying approximation schemes relevant for studies of particular aspects of the dynamics.  Attempts to enlighten the non-perturbative structure of QCD using many-body techniques have recently been undertaken by several authors \cite{adam-gluon,adam-conf} and  even the confinement scenario has been realized
within the Coulomb gauge\cite{adam-conf}.

Though much progress has been made, schematic models are still
very useful to shed some light onto the non-perturbative structure
of QCD. For example, in \cite{gluon99} the gluon sector was
investigated restricting the quark-gluon dynamics to an effective Hamiltonian for a fixed number of modes. The gluon spectrum was adjusted to reproduce lattice gauge calculations \cite{bali,peardon} and several other states have been predicted and
confirmed by lattice gauge calculations. In \cite{qcd-model1} a Lipkin type model was introduced where the fermion sector consisted of two levels, one at positive and the
other at negative energy, and a coupling to a boson level,
occupied by color-spin zero gluon pairs, was considered. Only
meson states were described. In \cite{qcd-model2} the model was
extended to include baryons. The nucleon resonances (especially
the Roper resonance) and $\Delta$ resonances were well described. One drawback of these model, however is that they are purely
 phenomenological and contain several parameters.
Our long term goal is to investigate classes of schematic models which are {\it derived} from QCD. For example, one expects the high energy quark and gluon modes be largely irrelevant in determining the structure of
the vacuum, and  lowest excitations, {\it e.g.} the pion or the
 $\rho$ meson states, color confinement, {\it etc.}
  One advantage of such schematic models is that {\it they are QCD} and will depend on only one or few well determined coupling constant(s). Preferably, such models should allow analytic or alternatively nearly analytic solutions, the latter requiring at most a numerical diagonalization.

Since it is a good practice to start from the simple cases first,
 here we present exact solutions to a
schematic model, which has all the most drastic approximations. The goal is to identify which of these play what role in the low energy dynamics.
 Hopefully by systematically relaxing these approximations an intuitive picture of QCD will emerge. The model presented here is derived from QCD, under the restriction to $SU(2)$ in color and flavor. Only quarks and antiquarks will be considered. The interaction via
gluons will be simulated via an effective, interaction reduced to the lowest
 momentum modes. These modes will correspond to the quarks and antiquarks restricted  to be in a single spacial orbital level
($S$-state). Apart from this, the model will depend only on one
parameter related to the energy of the color excitations thus
 ultimately irrelevant. 

The paper is organized as follows. In section II the derivation of
the schematic model, starting from QCD, is given and the appropriate particle content is identified. In section III the eigenvalues and the
basis states will be constructed and the physical states are
investigated in section IV.
In section
V
conclusions will be drawn and future developments
discussed.

\section{Derivation of the Model Hamiltonian}

As discussed above, by choosing an appropriate gauge a set of degrees of
freedom can be selected which
appears most natural for
 description of certain features of QCD.
In this case physical (gauge independent) quantities may be simpler to calculate  when the "correct" gauge is chosen. For example, to compute various deep inelastic amplitudes it is advantageous to formulate QCD in the light-cone gauge, while to compute low energy spectra Coulomb gauge seems to be the natural choice.
The Coulomb gauge has been extensively studied in
~\cite{adam-conf,TDLee}.
The Gauss's law can be used to eliminate the longitudinal component of
the electric field which leaves only the transverse gluons
representing
generalized coordinates and their conjugated
 momenta (given by the transverse electric fields). Schematically the
 Coulomb gauge Hamiltonian has the following structure
~\cite{adam-conf,TDLee},
 \begin{equation}
 H = K_q  + K_g + V_{qqg} + V_{g^3} + V_{g^4} + V_C. \label{ham}
 \end{equation}
 Here $K_q$ and $K_g$ are the kinetic energies of the quarks -antiquarks
 and gluons, respectively, and are  given by the Dirac and 
  Yang-Mills Hamiltonians. The next three terms have polynomial
 dependence on the canonical degrees of freedom and represent the
 local (anti) quark -gluon interaction, triple- and quartic- gluon
 coupling, respectively. Finally $V_C$ is the non-abelian
 generalization of the Coulomb potential.  In an abelian case, $V_C =
 \alpha \int d\x d\y \rho(\x) |\x - \y|^{-1} \rho(\y)$ represents the
 Coulomb energy between matter charges, which are described by the
 charge density $\rho(\x)$. For simplicity we have already dropped the
  Faddeev-Popov determinant, which as shown
 in~\cite{Szczepaniak:2003ve} can be accounted for by redefining the gluon wave
 functional.  

In a non-abelian theory like QCD the Coulomb potential depends not only on the relative separation between charges but also on the distribution of the gauge fields around them,
 \begin{equation}
 V_C = -g^2 \int d\x d\y \rho^a(\x) \left[ {\frac{1}{ 1 - \lambda^{\dag} }}
 {\frac{1}{ {\bf \nabla}^2 } } {\frac{1}{1 - \lambda}}
  \right]_{a\x;b\y}
 \rho^b(\y) .
 \end{equation}
The matrix elements of $1-\lambda$ are given by
 \begin{equation}
 [ 1 - \lambda]_{a\x;b\y}  = \delta_{ab} \delta(\x - \y)
   - g  f_{acb}  {\bf A}^c(\x) \cdot \nabla \delta(\x-\y) ,
         \end{equation}
 and the color-charge density is given by $\rho^a(x) = \psi(\x) T^a \psi(\x)
- f_{abc}{\bf A}^b(\x) \cdot {\bf E}^c(\x)$. ${\bf A}^a$, $-{\bf E}^a$
represent $a=1,\cdots N^2_c-1$ transverse gluon coordinates and conjugate
momenta, respectively and $T^a= T^a_{ij}, i,j = 1\cdots N_C$,
and $f_{abc}$ are the generators of the fundamental and
adjoint
representations of the color $SU(N_c)$ group. Thus, unlike QED, in QCD
to define the  potential between a state containing matter
(quark, antiquark) sources  it is necessary to know the gluon wave
functional of the state.  It was shown in ~\cite{var-conf} using a
variational ansatz for the gluon wave functional of the vacuum that $V_C$
leads to a confining interaction between matter sources.
Such an attractive interaction destabilizes the vacuum and leads to
formation of quark-antiquark condensates and chiral symmetry breaking.
The underlying mechanism
is
analogous to BCS superconductivity.

In the following we want to investigate the minimal
requirements, {\it e.g.} the minimal number of degrees of freedom in a
schematic model which yields the
pattern
of chiral symmetry breaking consistent with that expected in  QCD. Since the necessary condition for the condensate is existence of an attractive interaction we remove the gluon degrees of freedom (for example by fixing the gluon wave functional)  and approximating  the Coulomb kernel by a contact potential between quark charge densities. Under such approximation the Coulomb gauge Hamiltonian of Eq.~(\ref{ham}) reduces to,

\begin{eqnarray}
H   & = &  \int d\x \psid(\x) \left [-i{\vec\alpha}\cdot{\vec \nabla} + \beta m_0
 \right] \psi(\x) \nonumber \\
  & + &  g \int d\x \rho^a(\x) \rho^a(\x),
\end{eqnarray}
with the color charge density originating from quarks only,
 $\rho^a(\x)  = \psid(\x)T^a\psi(\x)$, and the coupling $g$
 which has mass dimension  $-2$ will be determined later.
  The quark fields $\psi(\x)$ represent $N_C \times N_f$ degrees of freedom. The generators of the flavor axial rotations are
 \begin{equation}
Q^\alpha_5 = \int d\x \psid(\x) \gamma_5 T^\alpha \psi(\x),
\end{equation}
 with $T^\alpha$ being the generators of flavor, $SU(N_f)$.
 In the limit of vanishing quark mass, $m_0 = 0$,
 the Hamiltonian is invariant under flavor-axial rotations,
\begin{equation}
\lim_{m_0 = 0} [Q^\alpha_5, H] = 0,
\end{equation}
while for a finite bare mass
\begin{equation}
[Q^\alpha_5,H] = -2 m_0 P^\alpha_5,
\end{equation}
with
\begin{equation}
P^\alpha_5 = \int d\x \psid(\x) \gamma^0 \gamma_5 T^\alpha\psi(\x).
\end{equation}
To obtain the particle content of the spectrum of this Hamiltonian we
first rewrite it in a basis
of massive quarks and anti-quarks defined by the operators
$b(cf\lambda\k)$ and $d(cf\lambda\k)$, respectively
with $c,f,\lambda,\k$ referring to color, flavor,
spin component and momentum, and related to the fields in the standard way

%%%new (indices c,f added to sum)
\begin{eqnarray}
\psi(\x)  & =  & \sum_{cf\lambda=\pm1/2}
\int {\frac{d\k}{(2\pi)^3}}
e^{i\x\cdot\k}   \left[ u(\lambda,\k) b(cf\lambda\k)  \right.
 \nonumber \\
  & + & \left.  v(\lambda,-\k) d^{\dag}(cf\lambda-\k) \right].
 \end{eqnarray}
%%%new-end
Here $u$ and $v$ are the eigenstates of the free Dirac Hamiltonian describing a fermion of mass $m$, which is not yet specified but is anticipated to be the constituent quark mass.  In terms of these quark operators the Hamiltonian is given by,
\begin{equation}
H = H_q + H_{q\bar q} + V.
\end{equation}
Here $H_q$ contains operators proportional to $b^{\dag} b$ and $d^{\dag} d$, $H_{q\bar q}$ contains pair creation and annihilation operators proportional to $b^{\dag} d^{\dag}$ and $d b$, and $V$ contains normal-ordered four-fermion operators.
 Since we are interested in studying the low energy phenomena we make the following simplification. First we confine quarks to a
 finite
%%%new%%% (better to use sphere, because orbital angular momentum should be a good quantum number)
%box
 box 
%%%new-end%%%
of volume ${\cal V}$. The momentum states become discrete,  with $\k \to \n$ and $\k =  2\pi \n /{\cal V}^{1/3}$, so that
\begin{equation}
 \int {\frac{d\k}{(2\pi)^3}} \to {\frac{1}{\cal V}} \sum_\n .
 \end{equation}
In the finite volume it is also useful to rescale the particle operators,
\begin{equation}
b(cf\lambda\k)  \to {\tilde b}(cf\lambda\n), \;\; b(cf\lambda\k)  =   {\cal V}^{1/2}{\tilde b}(cf\lambda\k),
\end{equation}
and the same of the anti-quark operator $d$. The new operator
 are dimensionless and satisfy
\begin{equation}
\{ {\tilde b}(cf\lambda\n), {\tilde b}^{\dag}(c'f'\lambda'\n') \} =
\delta_{cc'} \delta_{ff'} \delta_{\lambda\lambda'}\delta_{\n\n'}.
\end{equation}
In the following we will rename ${\tilde b},{\tilde d}$ back as $b$ and $d$,  respectively.
The final approximation is to retain only the lowest momentum states, {\it e.g.} $\n=0$. Thus from now on we drop the momentum index on the quark  operators. The next level of approximations  would include the  $P$- and higher waves. Within this approximation the Hamiltonian becomes,

\begin{eqnarray}
H  & = &  \sum_{1} ({\cal E} + m_0)  b^{\dag}_1 b_1 + \sum_{1} ({\cal E} + m_0)  d^{\dag}_1 d_1 \nonumber \\
& - & \sum_{1234} V_{qq}(1234) b^{\dag}_1  b^{\dag}_2 b_3 b_4
 - \sum_{1234} V_{{\bar q}{\bar q}}(1234) d^{\dag}_1  d^{\dag}_2 d_3 d_4 \nonumber \\
 & - &  2 \sum_{1234} V_{q{\bar q}}(1234) b^{\dag}_1 d^{\dag}_2 d_3 b_4.
\end{eqnarray}
with

\begin{equation}
{\cal E} =
 {\frac{g C_F }{\cal V}} \sum_{{\bf n}}^{{\bf n}_{max} } \delta_{{\bf n}0} =
 {\frac{g C_F  }{\cal V}}, \label{ecal}
 \end{equation}
  and
\begin{eqnarray}
V_{qq}(1234) & = & { \frac{g}{\cal V}} T^a_{c_1c_3} T^a_{c_2c_4}
\left[ \delta_{f_1f_3}\delta_{\lambda_1\lambda_3}\right]
\left[ \delta_{f_2f_4} \delta_{\lambda_2\lambda_4} \right]
\nonumber \\
V_{{\bar q}{\bar q}}(1234) & = &
{ \frac{g}{\cal V}} T^a_{c_3c_1} T^a_{c_4c_2}
\left[ \delta_{f_1f_3}\delta_{\lambda_1\lambda_3}\right]
\left[ \delta_{f_2f_4} \delta_{\lambda_2\lambda_4} \right]
 \nonumber \\
W_{q{\bar q}}(1234) &=&  { \frac{g}{\cal V}} T^a_{c_1c_4} T^a_{c_3c_2}
\left[ \delta_{f_1f_4}\delta_{\lambda_1\lambda_4}\right]
\left[ \delta_{f_2f_3} \delta_{\lambda_2\lambda_3} \right].
\nonumber \\
 \end{eqnarray}
Here $1=(f_1,c_1,\lambda_1)$ {\it etc.} denote all remaining
(discrete)
quantum numbers  of the particle labeled by $1$; $c_1$ denotes color, $f_1$ flavor and $\lambda_1$ spin projection.
It is worth noting at this point that with $S$-orbitals only the pair creation part of the Hamiltonian vanishes. Scalar quark-antiquark pairs
have quarks in relative spin-one coupled to one unit of orbital angular momentum which vanishes for $S$-waves.  Within these approximations the flavor axial charge
generators become
\begin{equation}
Q_5^\alpha = \sum_{12} \left( b^{\dag}_1 Q^\alpha_{12} d^{\dag}_2
 +d_1Q^\alpha_{12} b_2  \right), \label{q5-1}
 \end{equation}
 with
\begin{equation}
Q^\alpha_{12} = T^\alpha_{f_1 f_2}\delta_{c_1 c_2}\delta_{\lambda_1\lambda_2},
\end{equation}
and the pseudo-scalar charges $P^\alpha_5$  become
\begin{equation}
P^\alpha_5 = \sum_{12} \left( b^{\dag}_1 Q^\alpha_{12} d^{\dag} _2 - d_1  Q^\alpha_{12} b_2 \right).
\end{equation}

We also note that
\begin{equation}
[P^a_5,H] = -2m_0 Q^a_5,
\end{equation}
is still satisfied. For completeness, the vector flavor charges $V^\alpha$,
\begin{equation}
V ^\alpha = \int d\x \psid(\x) T^\alpha \psi(\x),
\end{equation}
become
\begin{equation}
V^\alpha = \sum_{12} \left( b^{\dag}_1 V^\alpha_{12}  b_2 - d^{\dag}_1 V^\alpha_{12} d_{2}  \right),
\end{equation}
with
\begin{equation}
V^\alpha_{12} = T^\alpha_{f_1f_2} \delta_{c_1c_2} \delta_{\lambda_1\lambda_2}.
\end{equation}

The Hamiltonian contains four  parts.  A non-interaction part,
quark-quark, and antiquark-antiquark potentials and a quark-antiquark
 potential.
We recall some basic properties of the particle operators.  In the following we concentrate on the case of two colors and two flavors.
Generalization
to arbitrary $N_C$ and $N_f$ is straightforward.
The creation and annihilation operators carry color, $c$, flavor, $f$  and
spin, $\lambda$ indices and these all range from
$-\frac{1}{2}$ to $+\frac{1}{2}$. We distinguish now between
co- and contravariant indices in order to denote the different
transformation properties of the fermion creation and annihilation
operators. We denote the creation and annihilation operators for quarks
 by $b^\dagger_\alpha$ and $b^\alpha$, respectively,
where $\alpha$ is a shorthand notation for $(cf\lambda)$.
Subsequently representation of
$SU(2)$-color, flavor, spin
will be similarly denoted by three numbers $(S_cS_fS)$, where
$S_c$ is the color angular momentum and similar for flavor, $S_f$, and  spin, $S$. Similarly  for the antiquark operators we have $d^{\dagger  \alpha}$ for the creation and
$d_\alpha$ for the annihilation operators.
 The anticommutation relations are now given by
$\left\{ b^\beta , b^{\dagger}_\alpha \right\} = \left\{ d_\beta , d^{\dagger  \alpha} \right\} = \delta^\alpha_\beta$.
The indices are lowered according to the following convention. If
 $a^\alpha$ denotes any of the four operators ($b^{\dag}$, $b$, $d$, or $d^{\dag}$) with an upper index, lowering this index corresponds to,
\begin{equation}
a^{cf\lambda} = (-1)^{\frac{1}{2} - c}
(-1)^{\frac{1}{2} - f} (-1)^{\frac{1}{2} - \lambda}
a_{-c-f-\lambda}.
\label{a-up}
\end{equation}
We can now rewrite the Hamiltonian. The non-interacting part is
trivial and given by,
\beqa
H_q & = & ({\cal E}+m_0) ( \hn_q + \hn_{\bar{q}}).
\label{h0}
\eeqa
with $\hn_q = b^{\dag}_\alpha b^\alpha$ and $\hn_{\bar q} = d^{\dag\alpha} d_\alpha$ being the quark and antiquark number operators,  respectively.
The quark-quark interaction is given by
\begin{eqnarray}
V_{qq} & = &  -\sum_{c'sf's\lambda 's} \frac{g}{{\cal V}} T^a_{c_1c_3}T^a_{c_2c_4}
\left[\delta_{f_1f_3}\right] \left[ \delta_{f_2f_4}\right] \nonumber \\
& \times &
{b}^\dagger_{c_1f_1\lambda_1}
{b}^\dagger_{c_2f_2\lambda_2}
{b}^{c_3f_3\lambda_3}
{b}^{c_4f_4\lambda_4}.
\label{hqq1}
\end{eqnarray}
Using
\beqa
T^a_{c_1c_3}T^a_{c_2c_4} = \frac{1}{2} \left( \delta_{c_1c_4}\delta_{c_3c_2}
-\frac{1}{2} \delta_{c_1c_3}\delta_{c_2c_4}\right) .
\label{tata}
\eeqa
and joining operators with common indices through the anticommutation  relations, we obtain in an intermediate step
\begin{eqnarray}
& & V_{qq}   =   -\frac{3g}{4{\cal V}} \hn_q - \frac{g}{4{\cal V}} \hn_q^2
 \nonumber \\
 & + & \frac{g}{2{\cal V}}\sum_{c_1c_2}
\left( \sum_{f_1\lambda_1}
b^\dagger_{c_1f_1\lambda_1} b^{c_2f_1\lambda_1} \right)
\left( \sum_{f_2\lambda_2}
 b^\dagger_{c_2f_2\lambda_2} b^{c_1f_2\lambda_2} \right).\nonumber \\
\label{hqq2}
\end{eqnarray}
Finally using Eq.~(\ref{a-up}) and coupling to definite
color, flavor and spin, we  arrive at
\begin{eqnarray}
V_{qq} & = & -\frac{3}{4}\frac{g}{{\cal V}} \hn_q
-\frac{2g}{{\cal V}} \sqrt{3} \left[\left[ b^\dagger \otimes
b \right]^{100} \otimes \left[ b^\dagger \otimes
b \right]^{100} \right]^{000}_{000}, \nonumber \\
\label{hqq3}
\end{eqnarray}
where $\left[ A^{\Gamma_1} \otimes B^{\Gamma_2}
\right]^{\Gamma}_{\mu}$ with $\Gamma = S_cS_fS$ and $\mu = cf\lambda$ denotes the coupling  of $A$ and $B$ in color, flavor and spin,
\begin{equation}
\left[ A^{\Gamma_1} \otimes B^{\Gamma_2}
\right]^{\Gamma}_{\mu} =
 \sum_{\mu_1\mu_2} \langle \Gamma_1 \mu_1, \Gamma_2\mu_2
  | \Gamma \mu\rangle
 A ^{\Gamma_1}_{\mu_1} B ^{\Gamma_2}_{\mu_2},
 \end{equation}
 and $\langle\Gamma_1 \mu_1, \Gamma_2\mu_2 | \Gamma \mu \rangle$ is the product of three Clebsch-Gordan coefficients
 in color, flavor and spin.

Note, that the quadratic dependence on the quark number operator
is canceled and only the linear dependence remains. The last term
in Eq.~(\ref{hqq3}) represents the color angular momentum squared, whose   components in spherical basis are given by,
\beqa
S^c_{q, m} & = &  \sqrt{2} \left[ b^\dagger \otimes b
\right]^{100}_{m00}
\nonumber \\
S^c_{\bar{q}, m} & = &  -\sqrt{2} \left[ d^\dagger \otimes d
\right]^{100}_{m00},
\label{col-ang}
\eeqa
for the quark and antiquark part, respectively. With this, the final form
of $V_{qq}$ is
\beqa
 V_{qq} & = & -\frac{3}{4} \frac{g}{{\cal V}} \hn_q +
\frac{g}{{\cal V}} \left( S^c_q \cdot S^c_q \right),
\label{hqq4}
\eeqa
 and we used $\left[ S^c_q \otimes S^c_q \right] =
 -\left(  {\bf S}^c_q \cdot {\bf S} ^c_q \right)$.
 In a complete analogy  one can show that the antiquark-antiquark part
is found to be,
\beqa
 V_{\bar{q}\bar{q}} & = & -\frac{3}{4} \frac{g}{{\cal V}} \hn_{\bar{q}} +
\frac{g}{{\cal V}} \left( {\bf S}^c_{\bar{q}} \cdot {\bf S}^c_{\bar{q}} \right).
\label{hqbqb1}
\eeqa
And finally for the quark-antiquark interaction is given by,
\beqa
V_{q\bar{q}} & = &
\frac{2g}{{\cal V}} \left( {\bf S}^c_{q} \cdot {\bf S}^c_{\bar{q}} \right),
\label{hqqb1}
\eeqa
Summing all terms  leaves us with a surprisingly simple Hamiltonian whose interactions are easily identified,
\beqa
 H  & = & \left( {\cal E} + m_0  - \frac{3}{4}\frac{g}{{\cal V}} \right)
\left( \hn_q + \hn_{\bar{q}} \right)
+ \frac{g}{{\cal V}} {\bf S}_c^2,
\label{hfin}
\eeqa
where  ${\bf S}_c^2 = \left( {\bf S}^c_q + {\bf S}^c_{\bar{q}} \right)^2$  is the  total color angular momentum squared.

\section{The Spectrum}

The basis used to diagonalize $H$ is determined by the number
of degrees of freedom each quark (antiquark) carries. There
are eight degrees of
freedom:
two spin times two flavor and times two color components.
The Fock space is thus finite and contains maximally eight quarks and eight antiquarks.
The group structure for each sector is given by \cite{hamermesh}
\beqa
U(8) \supset & U_c(2) \otimes & U_{fS}(4)  \nonumber \\
\left[1^{n_q}\right] & \left[h_1h_2\right] & \left[2^{h_2}1^{h_1-h_2}\right] \nonumber \\
& & \nonumber \\
U_{fS}(4) \supset & U_f(2) \otimes & U_S(2) \nonumber \\
\left[2^{h_2}1^{h_1-h_2}\right] & S_f & S ,
\label{gr-str}
\eeqa
The notation $\left[p_1p_2...p_n\right]$ refers to the Young
diagrams \cite{hamermesh}, which describes the symmetry under
permutation of a given irreducible representation (irrep) of
a unitary group.
In  Eq.~(\ref{gr-str})  we have $h_1+h_2=n_q$ and the reduction of the
flavor-spin group $U_{fS}(4)$ is given in \cite{hamermesh}. In
Table~\ref{table1} we give a list of the color-flavor-spin content
as a result of Eq.~(\ref{gr-str}).

\begin{table}
\begin{tabular}{|c|c|c|c|}
\hline
$n_q$ & $[h_1h_2]$ & $S_c$ & $\sum{(S_f,S)}$   \\
\hline
\hline
0 & $[0]$ & $0$ & (0,0) \nonumber \\
1 & $[1]$ & $\frac{1}{2}$ & $(\frac{1}{2},\frac{1}{2})$ \nonumber \\
\hline
2 & $[2]$ & $0$ & (0,0)+(1,1) \nonumber \\
2 & $[1^2]$ & $1$ & (1,0)+(0,1) \nonumber \\
\hline
3 & $[21]$ & $\frac{1}{2}$ & $(\frac{1}{2},\frac{1}{2}) +
(\frac{3}{2},\frac{1}{2})+(\frac{1}{2},\frac{3}{2})$ \nonumber \\
3 & $[1^3]$ & $\frac{3}{2}$ & $(\frac{1}{2},\frac{1}{2})$ \nonumber \\
\hline
4 & $[2^2]$ & $0$ & (0,0)+(1,1)+(2,0)+(0,2) \nonumber \\
4 & $[21^2]$ & $1$ & (1,0)+(0,1)+(1,1) \nonumber \\
4 & $[1^4]$ & $2$ & (0,0) \nonumber \\
\hline
5 & $[2^21]$ & $\frac{1}{2}$ & $(\frac{1}{2},\frac{1}{2}) +
(\frac{3}{2},\frac{1}{2}) + (\frac{1}{2},\frac{3}{2})$ \nonumber \\
5 & $[21^3]$ & $\frac{3}{2}$ & $(\frac{1}{2},\frac{1}{2})$ \nonumber \\
\hline
6 & $[2^3]$ & $0$ & (0,0)+(1,1) \nonumber \\
6 & $[2^21^2]$ & $1$ & (1,0)+(0,1) \nonumber \\
\hline
7 & $[2^31]$ & $\frac{1}{2}$ & $(\frac{1}{2},\frac{1}{2})$ \nonumber \\
\hline
8 & $[2^4]$ & $0$ & (0,0) \nonumber \\
\hline
\end{tabular}
\caption{ Color-flavor-spin content as a function on the number
$n_q$ of quarks. The list is equivalent for the antiquarks. }
\label{table1}
\end{table}

We now consider meson-like excitations, {\it i.e.} the Fock sector with equal number of quarks and antiquarks.
 For this case, the basis can be labeled by the following set of quantum numbers,
\begin{eqnarray}
& &  \mid n_{\bar{q}}=n_q; (S^c_{\bar{q}},S^c_q)S^cm^c;
(S^f_{\bar{q}},S^f_q)S^fm^f; (S_{\bar{q}},S_q)Sm \rangle, \nonumber \\
\label{basis} 
\end{eqnarray}
where $m^c$, $m^f$ and $m$ refer to the
magnetic color, flavor and spin projection. The eigenvalue of the Hamiltonian with respect to such states is given by
\beqa
E & = & \left( {\cal E} + m_0 - \frac{3}{4}\frac{g}{{\cal V}} \right)(n_q +n_{\bar{q}})
+ \frac{g}{{\cal V}}S_c(S_c+1). \nonumber \\
\label{energy}
\eeqa
For physical states with no net color only
the
first term contributes, and using
 Eq.~(\ref{ecal}) we find $E=m_0$ and the spectrum is degenerate with respect to flavor and spin. Color excitations are separated by a finite energy gap which is an artifact of the contact approximation for the quark interactions.  In full  QCD the splitting is expected to be infinite
  as the potential between the quarks grows with the relative separation.
 Nevertheless, one can investigate the
structure of colored excitations in the model, which might play a role in models like the quark-gluon glass condensate \cite{glass-qg}, important at high densities.

The energy solutions are simple and degenerate for
all
states with the same color.
%%%new (text changed)
At a first glance one might think that
%However,
the physical states should be
a certain sum over all basis states (Eq.~(\ref{basis})) with the same
color. As a consequence, in our schematic model one
%has to
would look
for adequate superposition of the degenerate states in order to
construct, {\it e.g.}, the physical vacuum state. One criterion used
%will
can
be to reproduce the quark condensate (see next section).
However, as we will show further below, arguments of coninuity require
that the lowest state has to be the vacuum state $|0\rangle$.
%%%new-end
To
get more physical solutions, it will be necessary to introduce an interaction
which lifts the large degeneracy of the Hamiltonian.

\section{Physical states and the chiral limit}

In the chiral limit, $m_0=0$ all color singlet states have zero energy.
The vacuum state should be identified as a state with all
scalar quantum numbers. Since the single quark-antiquark pair
in the $S$-wave
has pseudoscalar quantum numbers,
the vacuum will be given by a superposition
of states with an even number of quark-antiquark pairs
 with total color, flavor and spin zero.  The most general (unnormalized)
vacuum state can be schematically written as
\begin{equation}
| {\bf z} \rangle = |0\rangle + \sum^4_{n=1} z_n \left( b^{\dag} b^{\dag} d^{\dag} d^{\dag}  \right)^n |0\rangle. \label{vall}
   \end{equation}
    Since in the chiral limit
  all $J^{PC}=0^{++}$ states are degenerate
  in this model we cannot distinguish between the true vacuum and, for example the $\sigma$ meson.  Thus we take for the vacuum
 a state given by the sum of the perturbative vacuum $| 0 \rangle$
   and the state with the lowest number (two)
 of the quark-antiquark pairs coupled to definite color, flavor and spin.
Each pair can be written in the following equivalent form

\beqa
| S_cS_fS \rangle & = &
\frac{1}{\sqrt{2}} \left[ \left[ b^\dagger \otimes b^\dagger
\right]^{S_cS_fS} \otimes \left[ d^\dagger \otimes d^\dagger
\right]^{S_cS_fS} \right]^{000}_{000} |0\rangle. \nonumber \\
\label{two-qqbar}
\eeqa
Because the two-quark state has to be antisymmetric (the same for the
antiquarks) the only allowed color, flavor and spin values are
$(S_cS_fS) = (000), (110), (101)$ and $(011)$.
The coupling of first two quarks and then two antiquarks to a total
color, flavor and spin zero can be re-expressed easily in terms of the
coupling of two quark-antiquark pairs as follows,
\beqa
& \left[ \left[ b^\dagger \otimes b^\dagger
\right]^{S_cS_fS} \otimes \left[ d^\dagger \otimes d^\dagger
\right]^{S_cS_fS} \right]^{000}_{000} =
 - \sum_{S_c^\prime S_f^\prime S^\prime}
 & \nonumber \\
&
 \left\{
  \begin{array}{ccc}
  \frac{1}{2} & \frac{1}{2} & S_c  \\
  \frac{1}{2} & \frac{1}{2} & S_c  \\
   S_c^\prime & S_c^\prime  & 0    \\
  \end{array}
           \right\}
 \left\{
  \begin{array}{ccc}
  \frac{1}{2} & \frac{1}{2} & S_f  \\
  \frac{1}{2} & \frac{1}{2} & S_f  \\
   S_f^\prime & S_f^\prime  & 0    \\
  \end{array}
           \right\}
 \left\{
  \begin{array}{ccc}
  \frac{1}{2} & \frac{1}{2} & S  \\
  \frac{1}{2} & \frac{1}{2} & S  \\
   S^\prime & S^\prime  & 0    \\
  \end{array}
           \right\} & \nonumber \\
& \times \left[ \left[ b^\dagger \otimes d^\dagger
\right]^{S_c^\prime S_f^\prime S^\prime } \otimes
\left[ b^\dagger \otimes d^\dagger
\right]^{S_c^\prime S_f^\prime S^\prime } \right]^{000}_{000} &,
\label{recoupl}
\eeqa
where the symbols $\left\{ ... \right\}$ refer to the usual 9-j symbols
\cite{edmonds}.
For the vacuum we thus take the normalized state in the form,

\beqa
| z_0 z_1 \rangle = \frac{1}{\sqrt{1+2\rho^2 }}
\left( | 0  \rangle  + \sum_{S_c} z_{S_c} \sum_{S_fS} | S_cS_fS \rangle  \right),
\label{trial-state}
\eeqa
with $|S_cS_fS\rangle$ given in Eq.~(\ref{two-qqbar}).
Here we assumed that due to the degeneracy of the states with the same
color, there is no dependence of the trial state parameters $z_{S_c},
S_c=0,1$ on flavor and spin. In general the $z$-values are complex
 and can be written as $z_{S_c} = \rho_{S_c}e^{i\phi_{S_c}}$,
with $\rho_0=\rho cos(\phi )$ and $\rho_1=\rho sin(\phi )$,
and  $\rho = |z_0|^2 + |z_1|^2$ being the total radius.
In such a vacuum expectation values of $\hn_q$ and $\hn_{\bar{q}}$, which determine
  the  quark condensate, are given by
\beqa
\langle z_0 z_1 | \hn_q| z_0 z_1 \rangle = \langle z_0 z_1 |
\hn_{\bar{q}}| z_0 z_1 \rangle  & = &
\frac{4 \rho^2 }{1 + 2\rho^2 }. \label{nqnqb}
\eeqa
In the limit $z_{S_c} = 0$, $\hn_q = \hn_{\bar{q}} = 0$
as expected for the perturbative vacuum. For large
values of $\rho$, each expectation value approaches 2, as it has to be,
because then the main contribution comes from the two
quark-antiquark pairs.
Using  Eq. ~(\ref{nqnqb}) it is possible to define the collective potential
 as the expectation value of the Hamiltonian, the result is
\begin{equation}
V(z_0, z_1) = \langle z_0,z_1| H | z_0, z_1 \rangle =
\left( {\cal E}  - \frac{3}{4}\frac{g}{{\cal V}} \right)
\frac{8\rho^2}{1+2\rho^2},
\label{vz}
\end{equation}
which corresponds near $\rho =0$ to a harmonic oscillator and the potential saturates for $\rho \rightarrow\infty$ at
$4\left( {\cal E} - \frac{3}{4}\frac{g}{V} \right)$.
The use of such trial states 
played primordial role in nuclear physics to help understand the structure
 of a complicated many body problem \cite{iba} and might be here also
 of  great value when a more sofisticated Hamiltonian is used. 
 Because, as we showed above, the factor which contains ${\cal E}$ is zero
one obtains a flat potential which reflects the complete degeneracy
of color zero states. As already mentioned, the $z$ parameters are complex, but the expectation value above depends only on the total radius $\rho$. This implies that
equipotential lines flow along constant $\rho$ with arbitrary
angles $\phi_{S_c}$ and $\sqrt{\rho_0^2 + \rho_1^2}=\rho$.

To further determine parameters of the vacuum one can consider the quark condensate
$\langle {\bar q} q \rangle = \langle\bar{u}u\rangle = \langle\bar{d}d\rangle =
\langle {\bar \psi}(0)\psi(0)\rangle/2 \sim -(225 \mbox{MeV})^3 \sim - 1
\mbox{fm}^{-3}$ \cite{reinders},
\begin{equation}
\langle {\bar q}  q \rangle = - {\frac{1}{\cal V}} \left[ N_C N_S
  - \frac{1}{2}\left(\langle \hn_q \rangle + \langle \hn_{\bar q}\rangle
\right)\right] = -\frac{4}{{\cal V}} \frac{1+\rho^2}{1+2\rho^2}.
\label{qqbcondensate}
  \end{equation}
One might be tempted to use this to determine $\rho$ for given volume
({\it e.g.} taking as a volume of sphere of radius of $0.8\mbox{ fm}$
would yield $\rho=0.67$). This is however not correct since there are
 further constraints from the
  spontaneous realization of chiral symmetry breaking.
 Away  from the chiral limit, $m_0 \ne 0$ each additional quark-antiquark
 pair raises the energy by $2m_0$. Thus, for $m_0 \ne 0$ the vacuum has
 to be given by the single state $|0\rangle$, so $\rho = 0$.
 If there is no phase transition at $m_0=0$ then for all $m_0$ the
 vacuum should be identified with the $|0\rangle$ state and $\rho=0$.
 The quark condensate is then entirely determined by the volume and the
 total number of degrees of freedom,
 \begin{equation}
 {\cal V} = -   {N_C N_S} \langle {\bar q} q \rangle^{-1}  = 2.7 \mbox{fm}^3
 \end{equation}
 As expected for spontaneous breaking the generators of chiral symmetry, Eq.~(\ref{q5-1}), which can be also written as
 \beqa
 Q^5_f & = & \frac{\sqrt{N_C N_f N_S}}{2}  \left(
 \left[ b^\dagger \otimes d^\dagger
 \right]^{010}_{0f0} +
 \left[ d \otimes b\right]^{010}_{0f0} \right),
 \label{q5}
 \eeqa
 do not annihilate the vacuum, instead they mix the vacuum with the single pion state,
 \begin{equation}
 \langle \pi,f' | Q^5_f |0 \rangle= \delta_{f'f} f_\pi m_\pi \label{fpi} {\cal V}
 \end{equation}
  with $f_\pi = 93\mbox{ MeV}$ being the pion decay constant.  Chiral symmetry, and relativistic normalization of  single particle states,
 \begin{equation}
 \langle  \pi, f' | \pi f\rangle = 2 m_\pi {\cal V}, \label{norm}
 \end{equation}
  implies that in the chiral limit $m_0 \to 0$, $m_\pi = O(m_0^2)$ and $f_\pi = O(1)$.   Since pion has $J^{PC}=0^{-+}$ quantum numbers and is generated by the axial rotation from the vacuum the most general (unnormalized) pion state
 is given by,
 \begin{equation}
 |\pi \rangle \sim b^{\dag} d^{\dag}\left[ |0\rangle  + \sum_{n=1,4}
  w_i \left(b^{\dag} b^{\dag} d^{\dag} d^{\dag} \right)^n |0\rangle \right]
 \end{equation}
 Mixing with the vacuum through the axial charge, as given by Eq.~(\ref{fpi}), constraints the quark-antiquark component to
 \begin{eqnarray}
%|\pi,f\rangle & = & 
% f_\pi m_\pi {\cal V} {\frac{2}{\sqrt{N_C N_f N_S}}} 
%[b^{\dag} \otimes d^{\dag}]^{010}_{0f0} |0\rangle  \nonumber \\
%& + & \sum_{n=1}^4
% w_i \left(b^{\dag} b^{\dag} d^{\dag} d^{\dag} \right)^n |0\rangle, \label{pi}
 |\pi,f\rangle & = &    f_\pi m_\pi {\cal V}
  {\frac{2}{\sqrt{N_C N_f N_S}}} [b^{\dag} \otimes
  d^{\dag}]^{010}_{0f0} \left( |0\rangle  \right. \nonumber \\
  & + & \left. \sum_{n=1}^4 
  w_i \left( b^{\dag} b^{\dag} d^{\dag} d^{\dag} \right)^n
   |0\rangle \right), \label{pi}
 \end{eqnarray}

However, all states in the expansion in Eq.~(\ref{pi}) are eigenstates of the Hamiltonian  with increasing  eigenvalues  and physical state cannot be given such a linear combination. We thus conclude that the single pion state should be identified with the valence component alone,
\begin{equation}
|\pi,f\rangle =  f_\pi m_\pi {\cal V} {\frac{2}{\sqrt{N_C N_f N_S}}}[b^{\dag} \otimes d^{\dag}]^{010}_{0f0} |0\rangle
  \end{equation}
With the pion mass related to the bare quark mass by $m_\pi = 2 m_0$.
The normalization condition of Eq.~(\ref{norm}) then leads to
\begin{eqnarray}
f_\pi & = & \sqrt{ \frac{N_C N_f N_S}{ 2 m_\pi {\cal V}}}  = \sqrt{ -
\frac{N_f \langle {\bar q} q \rangle}{2 m_\pi } }  
 = 200 \sqrt{N_f}  \mbox{ MeV}. \nonumber 
\end{eqnarray}
The identification of other physical states with
the spectrum given in Table~1 is now straightforward.
Since the number of quarks and antiquarks are well defined and each additional
(anti)quark raises energy by $m_0$
the spectrum of single meson and baryon
states would correspond to stated with
a single $q{\bar q}$ pair and three quarks respectively.
States with other numbers of quarks or antiquark should be
identified with multi-particle states {\it e.g.}
$qqqq{\bar q}$ with a meson-baryon state.
Colored states are split from the physical color singlet states by
$g S_c (S_c + 1)/{\cal V}$ where $S_c$ is a half-integer or
integer total color for an
%%%new
%evan or odd
odd or even
%%%new-end
number of quarks and antiquarks in the state,
respectively, and $g$ is the effective strength of the colored interactions.
We thus see that it is now possible to take the limit $g \to \infty$
which is expected for the zero-mode component of a confining interactions
without affecting the physical spectrum.

\section{Summary}

Models play an important role in understanding complicated dynamical
%%%new%%%
structures.
%%%new-end%%%
Our goal here was not to build the most sophisticated model of low
energy QCD, but on the contrary to identify the most basis starting point for such an endeavor. Starting from the underlying QCD interactions
in the Coulomb gauge we have defined an approximations scheme which gave us a model for the interactions of the quark zero modes. The model is exactly solvable and physical states can be identified with help of  the symmetry patterns observed in the physical spectrum. In particular spontaneous breaking of chiral symmetry
enables to identify the vacuum state and the single pion state and conservation of the particle number by our model Hamiltonian then
%%%new
%lads
leads
%%%new-end
to mapping between the representations of the underlying $U(N_C \times N_f \times N_S)$ symmetry and the physical states.  We worked with the $N_C=2$ number of colors, but extension to $N_C=3$ is straightforward since the coupling and recoupling methods in
$SU(2)$ can be readily extended to $SU(3)$ (see, for example the appendix of Ref. ~\cite{jutta}).
 For the basis, instead of $U(8)$ we would start from $U(12)$ if flavor is still $SU(2)$ or $U(18)$ if flavor is also $SU(3)$. The reductions are known (see Ref. \cite{qcd-model1,qcd-model2,ramon}).
 In the model we find splitting between  physical states to be 
 proportional to the total bare mass of the quarks and anti-quars 
 independently of the strength of the color or confining interaction. 
 The color interaction is then responsible for lifting the
energy of the color non-singlet states. Even though the model respects
the pattern of chiral symmetry breaking the chiral behavior of the
physical constants, {\it e.g.} the pion mass and the pion decay
constant is not as expected. The pion mass turns out to be a linear
and not quadratic function of the symmetry breaking parameter, $m_0$,
and the decay constant depends on $m_0$. This is an
 expected behavior for large values of $m_0$, or the nonrelativistic
 quark model.
It is not surprising that our
schematic model away from the exact chiral limit of $m_0=0$ immediately
follows the pattern of a heavy quark theory since the model conserves
the quark number. This in turn is the consequence of reduction of the
quark degrees of freedom. With the gauge degrees of freedom
integrated out and quark Fock space reduced to the zero modes
there are no pair production interactions in the Coulomb gauge.
This suggests that by extending the Fock space to include a
limited number of non-zero momentum modes and/or addinggluon degrees
of freedom it may be possible to address the low energy
phenomena in a model with a finite number of degrees of freedom.

\section*{Acknowledgments}
 This work belongs to the DGAPA project IN119002. It was partially supported
by the National Research Councils of Mexico (CONACYT) and the US Department of Energy grant
 DE-FG0287ER40365.
 P.O.~Hess acknowledges the support of the Nuclear Theory Center at Indiana University where this work was initiated.

\end{document}